# Microring resonator as a Rayleigh mirror for broadband laser-cavity comb generation

**Aram A. Mkrtchyan[1,2*], Anastasia S. Netrusova[1], Mikhail S. Mishevsky[1,2], Zohran Ali[1], Nikita Yu. Dmitriev[3], Kirill N. Minkov[3], Dmitry A. Chermoshentsev[3,4], Yating Wan[2], Albert G. Nasibulin[1], Igor A. Bilenko[3] and Yuriy G. Gladush[1]**

[1]Skolkovo Institute of Science and Technology, 30 Bolshoy Boulevard, building 1, Moscow, 121205, Russia

[2]King Abdullah University of Science and Technology (KAUST), Thuwal 23955, Makkah Region, Saudi Arabia

[3]Russian Quantum Center, 30 Bolshoy Boulevard, building 1, Moscow, 121205, Russia

[4]Moscow Institute of Physics and Technology, (National Research University), Dolgoprudny 141700, Russia

These authors contributed equally: Aram A. Mkrtchyan, Anastasia S. Netrusova

E-mail: * aram.mkrtchyan@kaust.edu.sa

High-quality microring resonators (MRRs) have proven to be promising sources of optical combs generated from continuous-wave radiation. In addition to the primary comb that propagates along with the pump, Rayleigh scattering creates a comb that travels in the opposite direction. Normally, the scattering is a very weak, however, in the high-quality-factor MRR the power transferred to the backward-propagating comb can be quite significant. We demonstrate that the backward-propagating comb can be used as a feedback source for a fiber laser, effectively creating a nonlinear mirror for the laser cavity. By assembling a simple laser cavity comprising only active fiber and two mirrors, one of which is an integrated MRR, we show a robust self-starting comb generation with width exceeding 500 nm. We confirm the universal character of this approach for other types of microresonators, including whispering gallery mode resonators, by launching self-starting laser cavity combs with the crystalline toroidal cavity, coupled with a tapered fiber. This method provides significant simplification for the filter-driven laser cavity soliton generation, especially when free-space coupling is applied.

## 1. Introduction

The advancements in optical frequency combs generation[1–4] using high-quality (high-Q) microring resonators (MRRs) have given rise to numerous breakthroughs in metrology[5–14], spectroscopy[15,16], communications[17–20], as well as in quantum computing [21], quantum data processing [22], quantum sources [23,24], etc. For many applications, it is crucial to generate coherent frequency combs that, in the time domain, form dissipative cavity solitons – solitary waves that self-balance nonlinear phase shift, dispersion, losses, and energy flow[25–30]. However, on-chip soliton generation still faces serious challenges for practical implementation. The traditional method of soliton generation involves catching the soliton "step" by scanning through the cavity resonance with a narrow-line laser. This procedure is nontrivial, prohibits self-starting[27,31–33] and suffers from low pump-to-comb conversion efficiency, which typically does not exceed 5% for bright solitons[30,34–36].

To overcome these limitations, several approaches have been suggested, which can be divided into those that modify the microcavity and those that couple the pump laser to the cavity through optical feedback[37]. Helgason *et al.* have demonstrated a technique to induce a controllable frequency shift in a selected cavity resonance by using two linearly coupled anomalous-dispersion microresonators, achieving conversion efficiency of 50%[38]. To address the self-starting challenge, Yu *et al.* proposed an edgeless photonic crystal resonator, which enabled spontaneous soliton formation in place of Turing patterns[39]. Another common but intricate technique is to pump the MRR with an electro-optically modulated continuous wave (CW) external laser at a precisely selected frequency. This method was successfully demonstrated with a lithium niobate platform in 2021[40]. A popular and effective approach is the self-injection locking technique, which provides self-starting soliton comb generation[11,41–45] with up to 40% CW to soliton comb conversion efficiency[41]. Briefly, the laser diode locks to the traveling wave in the microcavity through the back-propagating wave emerging from Rayleigh scattering in the cavity[46]. This approach, besides cavity-soliton generation, is commonly used for narrow linewidth, low noise, and frequency-stabilized laser generation[47–50].

Another elegant technique that simultaneously solves both problems of low energy conversion and self-starting is the integration of a microcavity into a fiber laser resonator, which was first demonstrated to achieve stable ultrafast mode-locking in 2012[51]. By merging the fiber laser and microcavity, Bao *et al.* demonstrated 50-nm-wide laser cavity-soliton combs with 75% power in the comb[52]. Recently Maxwell *et al.* demonstrated that this system can exhibit self-starting soliton generation, which is naturally robust to the system perturbations and can spontaneously recover after disruption, which was explained by interplay between slow erbium fiber gain saturation and cavity thermal shift[53]. In these works, the authors used a four-port ring resonator inside the fiber laser ring cavity. Emission from fiber laser propagated through one bus waveguide to the microcavity and returned from the drop-port to the laser cavity through the other bus waveguide.

In this work, we further develop the technique of laser cavity soliton generation by demonstrating that a high-Q microring resonator can work as a nonlinear reflective mirror of the fiber laser cavity. Indeed, it has been shown that weak Rayleigh scattering into the mode of the resonator can be enhanced by an analog of the Purcell factor, leading to the formation of a back-propagating wave in the cavity and, under certain conditions, convert a major part of the pump power into it [54,55]. The spontaneous back propagating wave formation was previously employed to enable self-injection locking where even a small reflected power is sufficient to provide optical feedback to a laser. In the present work we demonstrate that the resonant back scattering from a MRR, working as a frequency selective mirror, can provide sufficiently large optical feedback to form a cavity for a fiber laser generation, leading to more than 500-nm-wide self-starting comb formation. This approach allows us to use a single bus waveguide MRR, which simplifies the fabrication and reduces losses compared to four-port cavity coupling, thereby maintaining the loaded Q-factor at higher level resulting in comb formation with better spectral properties. To prove the general applicability of this approach, in addition to integrated MRR, we also demonstrate the self-starting comb generation for a crystalline $MgF_2$ microcavity coupled with tapered fiber. This approach can be very efficient for robust laser cavity comb generation with microcavities that requires coupling from free space through prism or tapered fibers or when MRR is combined on the same photonic circuit with integrated amplifier[56–59].

## 2. Results

The backward-propagating wave formation is a complex process defined by the relation between microring resonator parameters, namely, intrinsic loss ($\kappa_0$), coupling rate ($\kappa_C$), and backward-wave coupling rate ($\gamma$)[55]. Our MRR has 1 THz free spectral range (FSR), nearly 1 million Q-factor, - 13.5 $fs^2$/mm second-order dispersion, and following relation between its parameters: $\kappa_0<\kappa_C<\gamma<\kappa_0+\kappa_C$ (see the characterization section for details). Under low pump power this relation guaranties prominent back propagating wave without significant mode splitting. When the power is increased enough for comb formation back-propagating comb also forms with a spectrum similar to forward propagating, which we observed under an external tunable CW laser pump (Figure S1).

We used the MRR to form a fiber laser cavity by edge-coupling the chip with the MRR to the polarization maintaining (PM) Er-doped fiber amplifier (EDFA) and introducing fully reflecting gold mirror on the other side (see Figure 1a). Additionally, the cavity includes a PM-WDM with a 980 nm laser diode and a pair of fiber connectors for simplifying coupling alignment. The total laser cavity length is 3.2 m, corresponding to a 32.5 MHz FSR. At the threshold pump power of 43 mW, laser generation starts at the wavelength of one of the resonances due to build up of the back propagating wave in the MRR. As the pump power increases, the optical comb grows, reaching its width of more than 500 nm at the maximum available pump power of 500 mW, corresponding to 91 mW in the fiber before the chip (Figure 1b). Remarkably, turning the laser off and on led to the immediate recovery of the comb (see Supporting Video). This behavior is similar to laser cavity solitons, demonstrated earlier with a four-port cavity nested into fiber laser, where the soliton was manifested as a stable attractor due to the interaction of slow gain saturation and MRR heating[53].

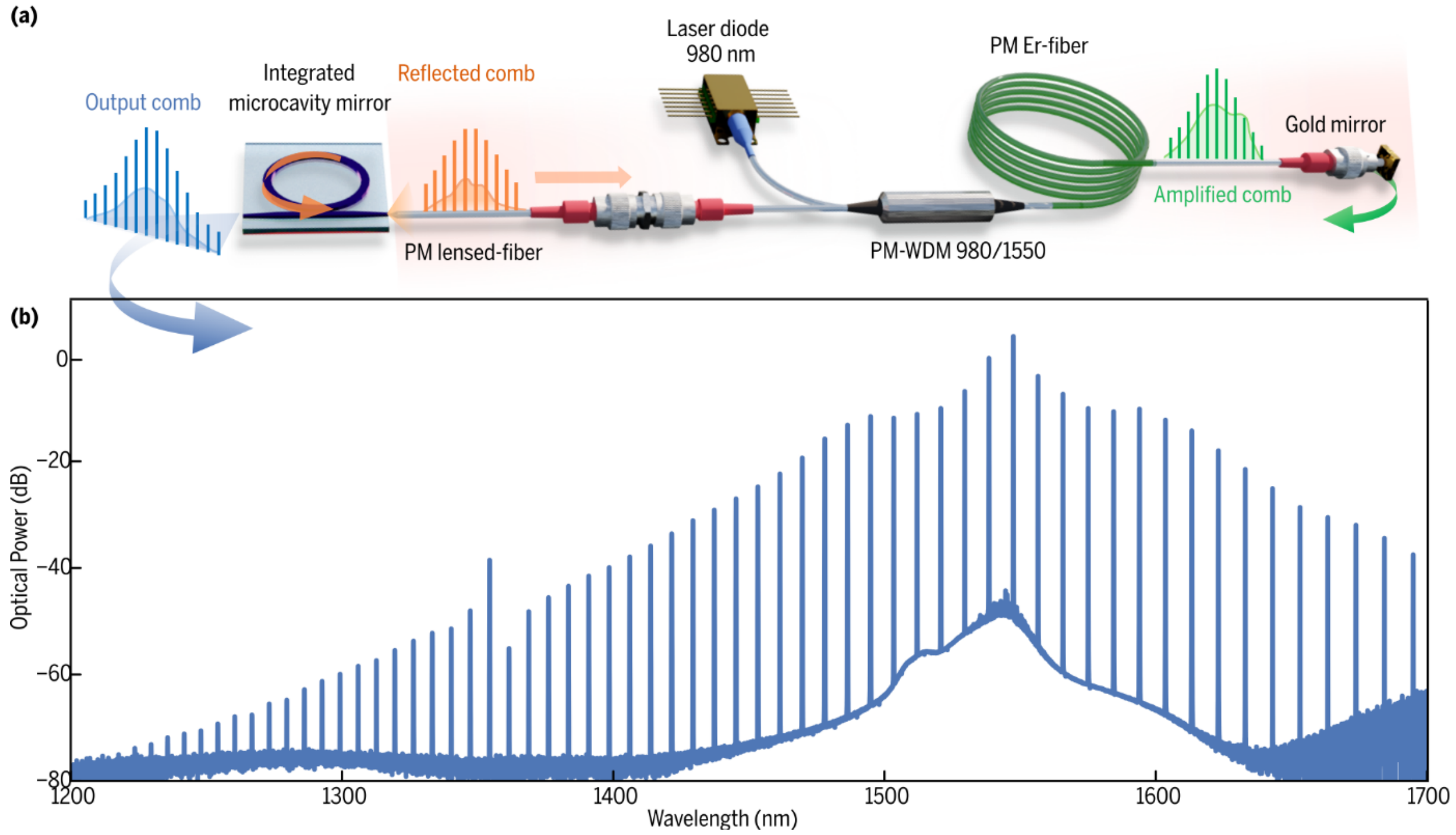

**Figure 1. Generation of broadband frequency combs using high-Q silicon nitride ($Si_3N_4$) microresonator nested into fiber laser cavity in reflection scheme.** a) Schematic illustration of the fiber laser with integrated photonic chip for cavity soliton generation: one-meter-long polarization-maintaining Er-fiber pumped with 980 nm laser diode through a broadband wavelength-division multiplexer 980/1550 nm (PM-WDM). All fibers in the system were polarization-maintained and exhibited a second order dispersion of -23 $fs^2$/mm. Lensed fibers are used to couple light to the chip. $Si_3N_4$ microring resonator has 1 THz repetition rate, nearly 1 million Q- factor and anomalous dispersion. b) Cavity soliton spectrum with width exceeding 500 nm at 112 mW intracavity power.

We note that our scheme does not include an optical filter and a delay line, which are usually presented in such systems[51–53,60,61]. We found that, in our case, these elements have a very small effect on the generation performance, and we eliminated them from the cavity. A brief discussion can be found in the Supporting Information (see section SI2).

To investigate the fiber intracavity dynamics of the comb, we introduced a 2x2 70/30 coupler into the fiber resonator (see Figure S7), which allows us to track the intracavity back-reflected and amplified combs. It introduces additional losses to the system, reducing the intracavity power, but the general behavior is preserved. The optical spectra for transmitted and reflected combs (Figure 2a, b) demonstrate slight differences while maintaining the triangular shape typical for soliton combs. The apparent reduction in the reflected comb width arises from additional losses at the coupler and circulator located on the way to a photodetector. Figure 2c shows the comb after double propagation through the erbium doped fiber, demonstrating significant amplification within the erbium gain spectrum region leading to comb width reduction). To perform autocorrelation measurements, external commercial amplifier was used (see Supplementary materials SI4). Autocorrelation traces clearly show pulse trains with 1 THz repetition rate for output, reflected and amplified combs (see Figures S4). However, significant distortion of the comb spectra by external amplifier does not allow to determine the pulse width. To address the pulse formation and its characteristics we modeled the intracavity dynamics using the coupled mean field equations for MRR forward propagating, backward propagating and fiber-amplified combs A detailed description of the model and numerical procedure is provided in the Supplementary Information (section SI3). The dashed envelopes on the Figures 2 (a-c) correspond to numerical simulation, which shows excellent agreement for output and amplified combs. The insets in Figure 2 (a-b) shows corresponding modeled pulse trains, which have bell on a pedestal shape for the intracavity pulses. The pedestal is reproduced if pulse shape is recovered from Fourier transform in assumption of full coherence (see Supplementary materials section 5). The simulated pulse in the fiber laser section, shown on the inset of Figure 2c, is much broader compared to MRR pulses due to spectral narrowing after the amplifier.

To further prove soliton comb operation, we checked whether comb lines had red-detuning using laser-scanning spectroscopy (LSS) technique[52,60]. We introduced external tunable 10 kHz linewidth CW laser emission into the fiber cavity through the fiber optic circulator attached to the additional 70/30 coupler (see section SI6 in Supporting Information). The CW laser swept through one of the MRR "hot" resonances simultaneously with comb generation. For the well-formed comb of Figure 2a, the beating modes (Figure 2d-f) clearly indicate the red-shift of the central and two side modes with respect to cavity resonance, indicating that the system may operate in the soliton regime[62,63]. Important to note that we always observe only single mode inside the resonator and never observe hopping between fiber resonator modes. We conclude that the laser operates on a single supermode, i.e. a sequence of equidistant laser modes with 1 THz separation, which is also evidence for pulsed generation with 1 THz repetition rate.

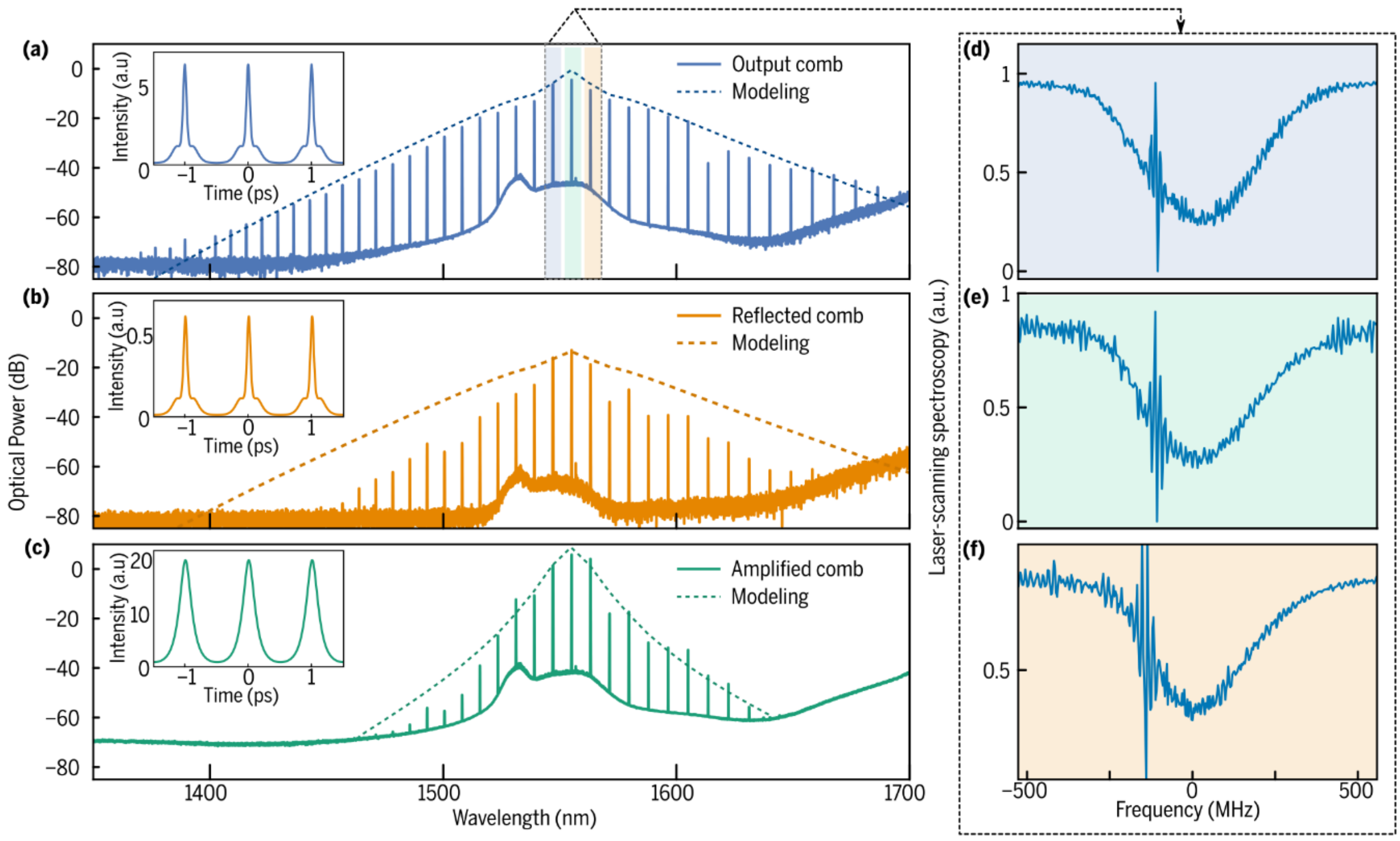

**Figure 2. Single soliton comb generation in hybrid-laser scheme**. a–c) Measured output (a), reflected (b), and amplified (c) comb spectra. Dashed curves denote the intracavity spectral envelopes computed using the mean-field equations. The insets show pulse trains obtained from the same model, since direct autocorrelation measurements are not feasible due to insufficient pulse energy; d-f) laser-scanning spectroscopy of the 1542 nm (d), 1554 nm (e), and 1562 nm (f) microcavity resonances measured for an output comb. All plots indicate red-detuned state for a central and side modes of the comb.

Next, we measure pump power-dependent spectrum, mode detuning, and fiber intracavity power simultaneously at the chip, as well as reflected and transmitted positions as indicated on Figure S7. The power entering the bus waveguide (green line in Figure 3a) shows a linear dependence on the pump power. This power is converted into forward-propagating comb (blue area) and backward-propagating comb (red area), while the rest of the power (green area) corresponds mainly to losses in the MRR. Additionally, we observe that the transmitted and reflected combs exhibit a nonlinear dependence on the pump power.

To investigate the comb formation in more detail, in Figure 3b we plot the power fractions of transmitted (blue line) and reflected (red line) combs, as well as corresponding mode detunings as a function of power inside the bus waveguide. The corresponding comb spectra are shown in Figure 3c. Below the lasing threshold, almost all the power is transmitted, while a small reflection of the order of 1% can be attributed to the reflection from the chip facet (see Figure S8 for the fractions of transmitted and reflected light measured in the fibers). As the power inside the waveguide approaches 1 mW, a rapid growth of reflection to 48% is observed, accompanied by the drop in the transmitted signal to 10%. LSS measurements show that at CW state the laser mode is blue-detuned with respect to the MRR resonance. However, as the power increases and the comb is excited, the laser supermode shifts to a red-detuned state. This behavior resembles the comb formation observed in the filter-driven four-wave mixing setup reported by Rowley *et al.*[53], with the notable difference that the multiple soliton formation is never observed for this MRR at the available pump power. Interestingly, as the comb develops, a significant drop in the reflected signal to 23% is observed, along with an increase in the transmitted comb power fraction to 36%. We attribute this behavior to the nonlinear interaction between forward and backward-propagating waves through the cross-phase modulation [64,65]. Additionally, the fraction of light lost in the MRR can be determined by summing the fractions of transmitted and reflected light (orange line in Figure 3b). Once the pump power is sufficient to form a CW laser line, the losses approach 40% and remain near this value independently of the pump power. The gradual change in losses resembles the shift of the laser supermode through the MRR resonances with the maximum losses corresponding to the supermode position at the resonance minimum.

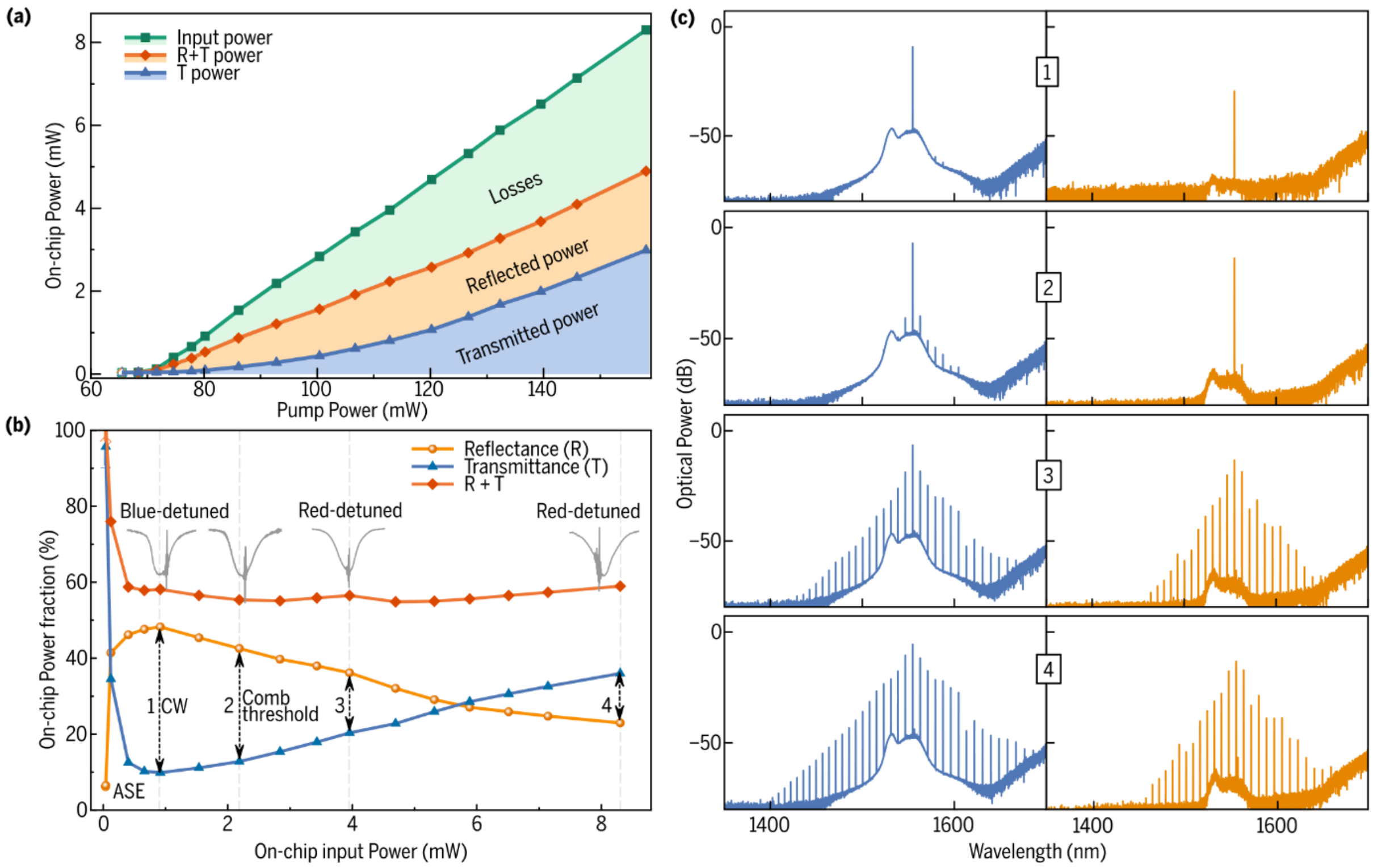

**Figure 3. Intracavity dynamic measurements.** (a) On-chip power distribution diagram for the transmitted and reflected combs, as well as the power propagated after a double EDFA pass (input), plotted as a function of pump power. (b) Chip reflection (red) and transmission (blue) as a function of the power in the bus waveguide. The brown line demonstrates the sum of transmitted and reflected signals, indicating nearly 40% losses in the chip. Numbers correspond to the spectra shown in (c) for different pump powers for transmitted (blue) and reflected (red) combs.

The reflection from the chip facet, reaching 1%, is sufficient to start CW laser generation even without the MRR. This raises the question whether it affects the generation starting dynamics and whether starting mechanism similar to self-injection locking could be responsible here[44]. To prove that comb generation self-starting can be achieved solely with a high-Q MRR mirror, we replaced the chip with a crystalline toroidal microresonator ($MgF_2$; $10^9$ Q-factor, 35 GHz FSR and normal dispersion) coupled through a tapered fiber (Figure 4a). The relatively thick taper with a 3 μm waist was prepared to ensure the parasitic reflection fraction is below $10^{-5}$. The response of the system to the pump power generally resembles the behavior observed with the integrated MRR. The reflected power from the toroidal resonator as a function of the power incident from the active fiber is shown in Figure 4b. When the incident power reaches 3 mW (86 mW pump power) the reflection rapidly changes to 0.1%, accompanied by the formation of CW laser emission (see insets in Figure 4b for spectra). Further increasing pump power leads to comb formation[49,66–68]. The system also demonstrates self-starting behavior, which resembles, even if the pump power is turned off and on. However, we did not manage to achieve a coherent comb state with this approach. Instead, the comb always consists of several subcombs corresponding to different families of toroidal cavity resonances[69]. A more detailed investigation of the generation regimes in a fiber laser with a nested crystalline resonator is left for future work.

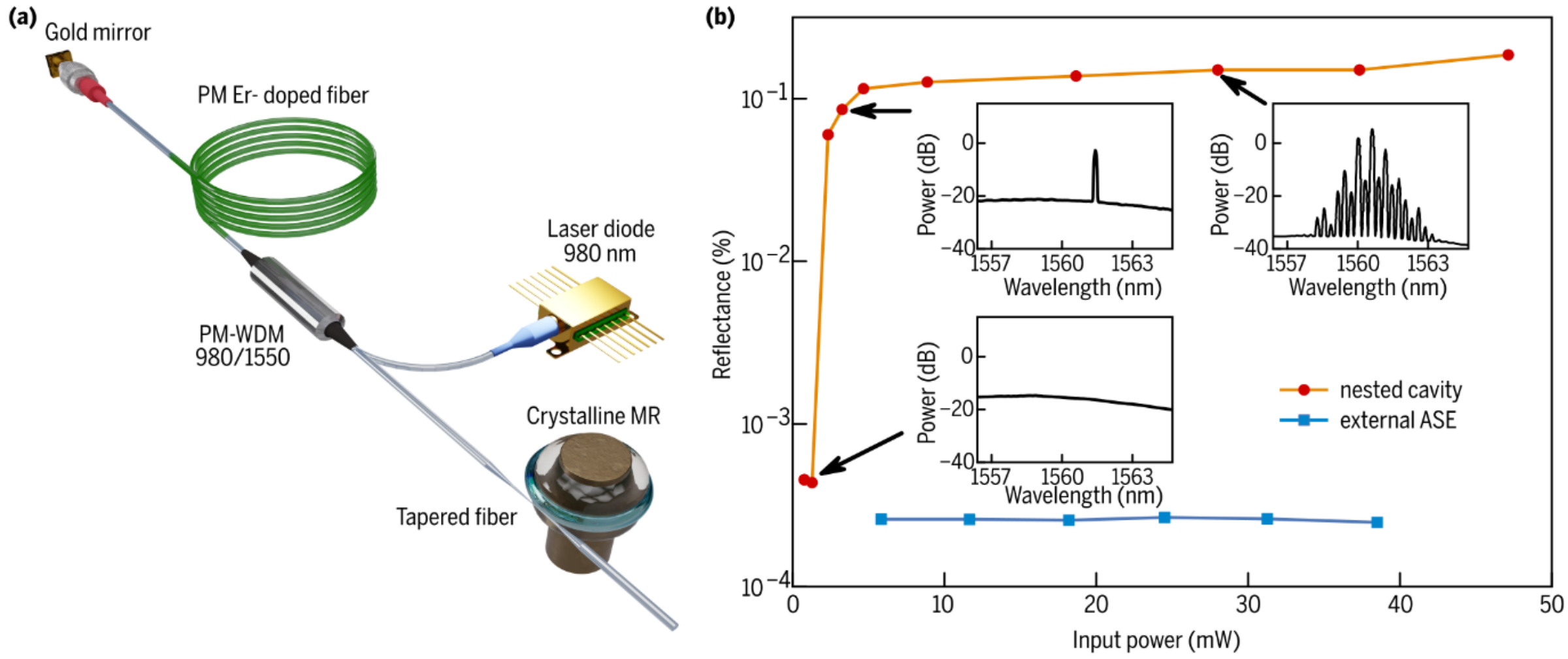

**Figure 4. Generation of frequency combs using high-Q $MgF_2$ crystalline toroidal microresonator nested into fiber cavity, where only backward light propagation due to Rayleigh scattering create feedback.** a) Hybrid-laser scheme with the crystalline $MgF_2$ MRR with $10^9$ Q-factor, 35 GHz FSR and normal dispersion, was used to generate whispering gallery modes. b) Reflection from crystalline microcavity as a function of power incident from the active fiber (red curve). The corresponding spectra are shown in the insets. Blue curve shows the reflection from an external amplified spontaneous emission (ASE) source, which is independent of the source power.

## 3. Conclusion

In conclusion, we have demonstrated the application of a ring microcavity as a partially reflecting mirror to form a laser cavity for efficient and robust comb generation. In our setup, the MRR serves two functions. First, it is responsible for the comb formation from the EDFA amplified light. Second, the back-propagating comb, spontaneously created by Rayleigh scattering, provides optical feedback for the laser cavity. We used two types of cavities - the silicon nitride integrated MRR with a single bus waveguide and the toroidal crystalline microresonator coupled by the tapered fiber. In both cases, we observed robust self-starting comb formation, which demonstrates full recovery under perturbations. For the integrated MRR, we show comb formation with 1 THz FSR and width of more than 500 nm. The coherence of the comb is supported by red-detuning of the main comb lines, autocorrelation measurements as well as results of numerical modeling. This comb shows long-term stability and recovers its state under external perturbations. We suggest that this method expands the applicability of the previously proposed laser cavity soliton generation method and provides a significant advantage when single coupling is desirable.

## 4. Methods

***Integrated MRR***: The photonic $Si_3N_4$ chip was fabricated in Ligentec, Switzerland, and contains structures consisting of 25-um radius microring resonators and bus waveguide (Figure 5a-b). All structures were fully characterized using a method based on a pre-calibrated Mach-Zehnder interferometer [55]. Characterization includes measurements of the microresonator transmission spectrum (Figure 5d), microresonator dispersion landscape (Figure 5e) and distributions of Q-factor, intrinsic loss $\kappa_0/2\pi$, coupling rate $\kappa_C/2\pi$ and backward-wave coupling rate (mode splitting) $\gamma/2\pi$ for all resonances in wavelength range from 1510 nm to 1620 nm (Figure 5f). For the central line measured parameters correspond to $\kappa_0/2\pi$= 95 MHz, $\kappa_C/2\pi$ = 116 MHz and $\gamma/2\pi$ = 142 MHz with the linewidth 270 MHz. According to the characterization results, the microring resonator exhibits anomalous group velocity dispersion and the pump mode corresponding to the central line of the generated microcomb features $Q \sim 0.9 \times 10^6$ (Figure 5c).

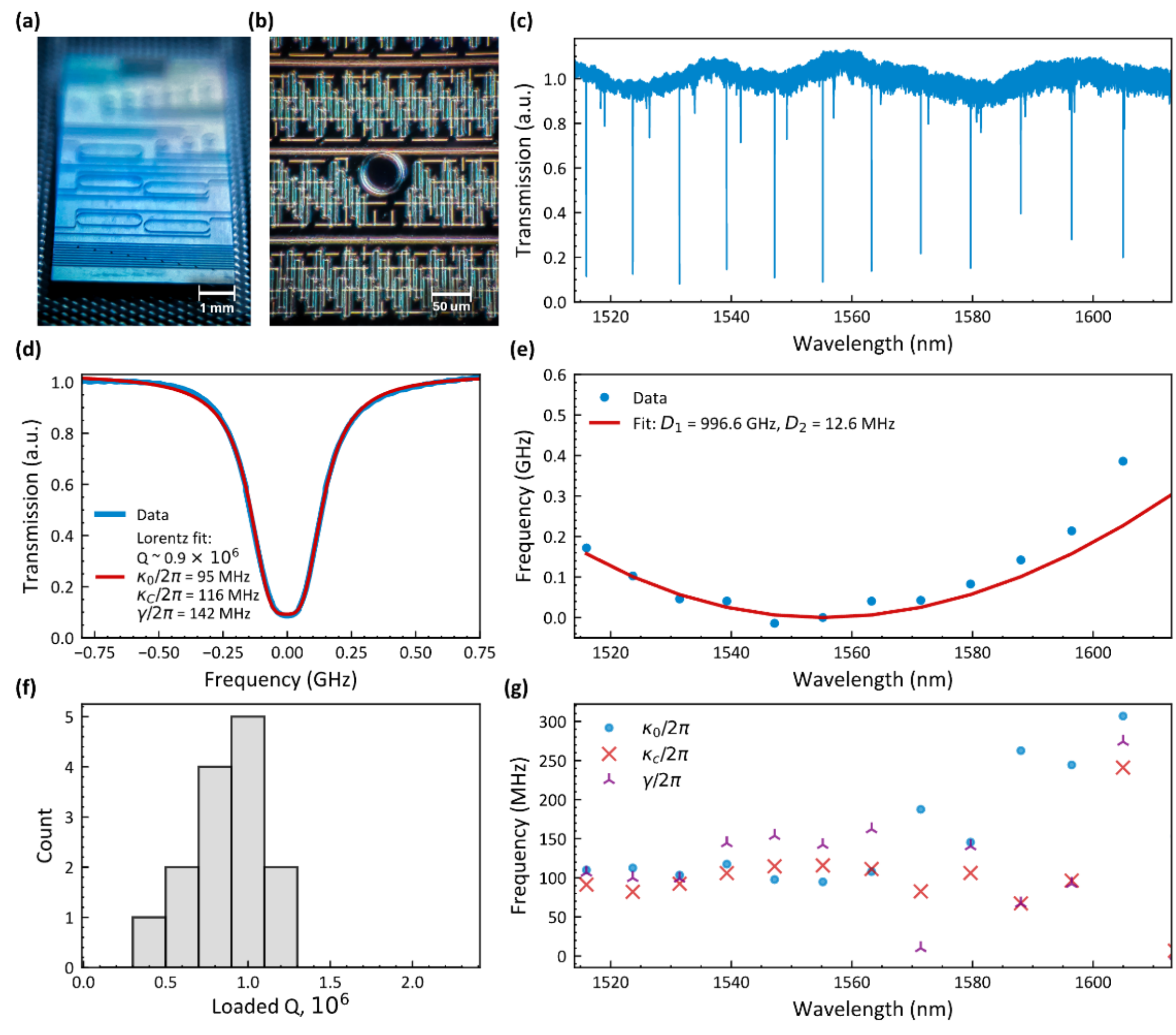

**Figure 5. Characterization of the $Si_3N_4$ microring resonator.** a) Photonic chip containing 1THz microring resonators. b) Structure under study consisting of the microring resonator and bus waveguide. c) Microresonator mode spectrum. d) Parameters of mode corresponding to the central comb line. e) Dispersion landscape (microresonator eigenfrequencies deviation from the uniform FSR-grid). f-g) Distribution histograms of the Q-factor, intrinsic loss ($\kappa_0/2\pi$), coupling rate ($\kappa_c/2\pi$), and backward-wave coupling rate (mode splitting, $\gamma/2\pi$) for microresonator modes in the range 1510–1620 nm, estimated using a Lorentzian function, with backscattering taken into account.

***Measurement Technique:*** Laser scanning spectroscopy and chip resonant/ out-of-resonance measurements were carried out by clean sweeping narrow-linewidth tunable Pure Photonics PPCL550 laser through the circulator and the 2.5 μm waist size lensed fiber (AMS Technologies, PM-lensed fiber

TPMJ-3A-1550-8/125-0.25-7-2.5-14-2-AR). For laser scanning spectroscopy, we used a Thorlabs DET08CFC/M 5 GHz photodetector and a Tektronix MDO3102 oscilloscope. Average laser power characteristics were measured by Thorlabs S132C power sensors. The spectral properties of optical frequency comb were measured by DEVISER AE8600 and Yokogawa AQ6370E spectrum analyzers with 600 – 1700 nm operating window. The temporal characteristics of solitons was investigated with a Femtochrome FR-103XL autocorrelator with 1 fs resolution.

***Fiber Laser:*** The fiber laser cavity shown in Figure 2a was assembled from the EDFA spliced to the gold mirror from on one side and edge-coupled to the MRR on the other side through the lensed fiber. All the fibers used in this work were polarization-maintaining (PM). To make the fully reflecting mirror, we sputtered gold onto the fiber ferrule, resulting in a 98% reflection coefficient. The EDFA consisted of 1-meter Liekke Er-doped active fiber with an 80 dB/m absorption coefficient at 1530 nm, pumped by 980 nm laser diode with a maximum power of 500 mW through a 1550/980 wavelength division multiplexer (1550/980 WDM). The total laser cavity length was 3.2 m, corresponding to 32.5 MHz FSR.

***Crystalline microresonator:*** Microresonator and tapered fiber for coupling were fabricated at the Russian Quantum Center. The resonator was manufactured from magnesium fluoride using diamond turning[70]. The diameter of the resonator is 1.94 mm, its thickness is 1.0 mm, and the radius of curvature is 0.35 mm. The taper was produced from a single-mode fiber (SMF28 Ultra) using a hydrogen flame[71].

**Supporting Information.** Supporting Information and Supporting Video are available online or from the author.

**Acknowledgements.** This work was supported by the Russian Science Foundation (RSF) under Grant No. 25-22-00388.

**Author contributions.** Y.G.G., I.A.B., A.A.M. developed the original research idea and designed the experiments. A.A.M., A.S.N., Z.A., N.Y.D. performed the experiment and analyzed the data for the integrated chip. A.S.N., M.S.M. developed the mean-field model and wrote the simulation code. K.N.M., M.S.M. performed the experiment and analyzed the data for crystalline resonator. A.A.M., A.S.N. drafted the manuscript, Y.G.G., I.A.B., D.A.C., and Y.W. contributed to manuscript writing. Y.G.G., A.G.N., I.A.B. supervised the research. All authors contributed to discussion of the results and experimental setup.

**Competing interests.** The authors declare no conflicts of interest.

**Data Availability Statement.** Data underlying the results presented in this paper are not publicly available at this time but may be obtained from the authors upon reasonable request.

# Supporting Information

## Microring resonator as a Rayleigh mirror for broadband laser-cavity comb generation

*Aram A. Mkrtchyan, Anastasiia S. Netrusova, Mikhail S. Mishevsky, Zohran Ali, Nikita Yu. Dmitriev, Kirill N. Minkov, Dmitry A. Chermoshentsev, Yating Wan, Albert G. Nasibulin, Igor A. Bilenko and Yuriy G. Gladush**

### 1. Forward- and backward-propagating combs with CW pump

We pumped the chip via an edge-coupled external narrow-linewidth tunable CW laser through a circulator and scanned across the MRR resonance, observing evolution of blue-detuned regimes from the primary to modulation instability (MI) comb (see transmitted comb in **Figure S1**). The backward-propagating comb had a shape of spectrum similar to the forward-propagating one and contained 1.8% of the incident power returning back to the fiber, corresponding to 8% on-chip reflection (Figure S1).

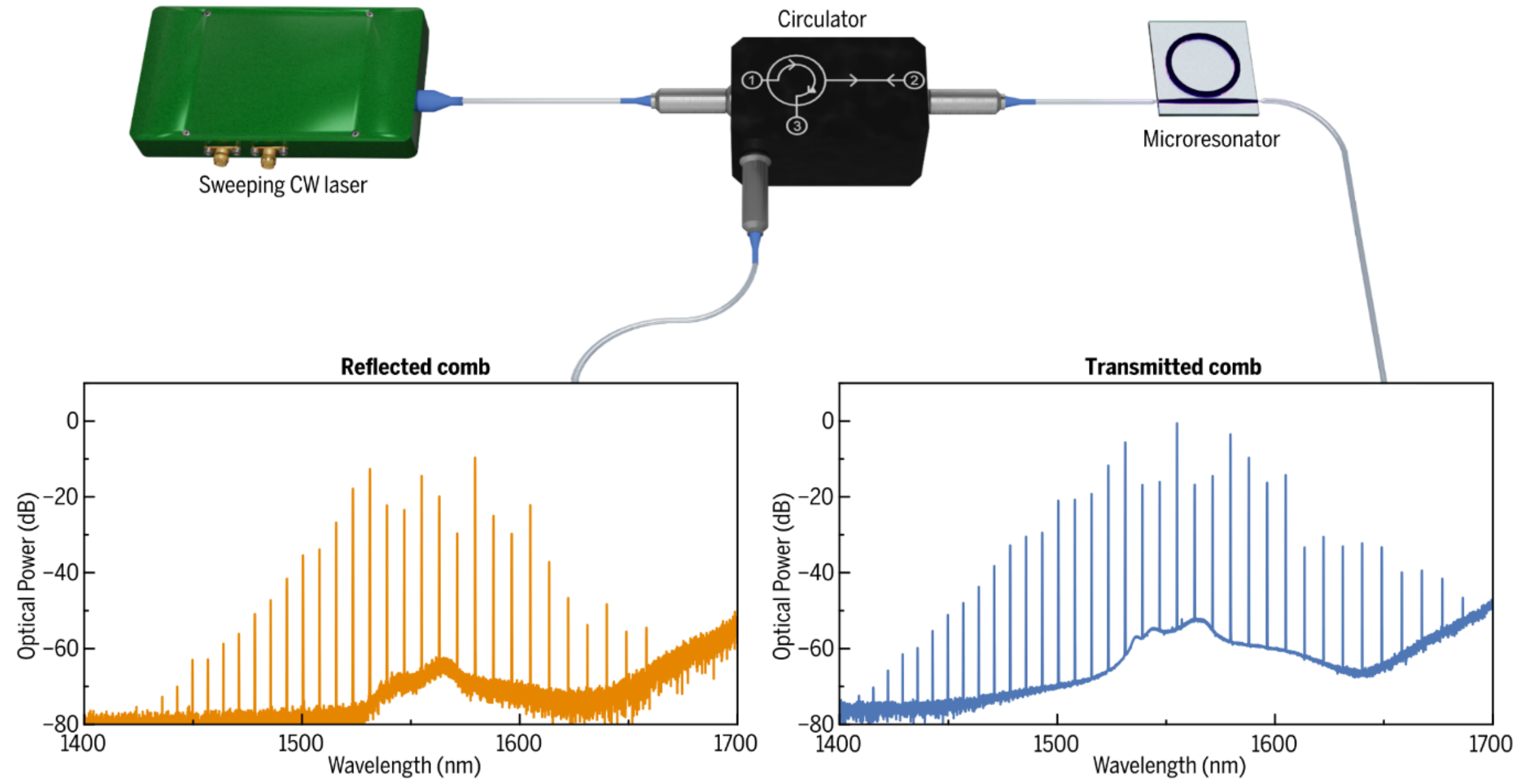

**Figure S6. Reflected and transmitted spectra of external CW laser-generated frequency combs in microcavity.** The setup includes the $Si_3N_4$ microresonator with the following parameters: 1 THz free spectral range (FSR), 1 million Q-factor, -600 $fs^2$/mm second-order dispersion (anomalous), and 0.55 bus waveguide-to-MRR coupling coefficient at 1550 nm. Tunable laser emission goes through polarization maintaining (PM) circulator followed by PM-lensed fibers butt-coupled to the chip to excite TE-mode of the microcavity.

### 2. Variations of hybrid-laser scheme

We have compared our approach for the linear laser scheme (**Figure S2a**) and in the ring cavity scheme where the MRR is coupled to the laser cavity through the circulator (see Figure S2c). Both schemes demonstrated comparable soliton self-starting performance. However, the linear scheme yielded more efficient comb generation (Figure S2b) in comparison to the ring (Figure S2d) cavity due to additional losses in the ring cavity caused by the double pass through the circulator.

Soliton generation requires the laser cavity to be an isnteger multiple of the microcavity length. To satisfy this demand, a free-space delay line is usually presented in the fiber resonator. We found that for our 25-um radius MRR, the required condition could be met by a slight change (within 3 μm) in the gap width between the lensed fiber probe and the chip edge. This variation was small enough not to lead to a significant decrease in the fiber-to-chip coupling efficiency. Therefore, we did not use an additional delay line in the scheme.

We tracked how comb generation changes with the presence of a tunable spectral filter inside the fiber ring resonator (see Figure S2e-g), as previously reported in the works [1–4]. The role of the filter is to ensure power flow from the main two modes to the wings to ensure coherent comb formation. In our case we found that if the filter transmits at least two comb lines, the comb is formed with little difference to the comb without filter (Figure S2e-f). We suggest that 1 THz FSR is large enough so that sufficient spectral filtering is provided by the gain bandwidth. We also observed that the stability of comb generation varies with the band-pass filter position and the gain bandwidth. Specifically, stable comb generation occurs when the band-pass transmission is red-shifted relative to the gain central wavelength (Figure S2e-f), while instability arises in the blue-shifted region (Figure S2g), consistent with previous findings[4].

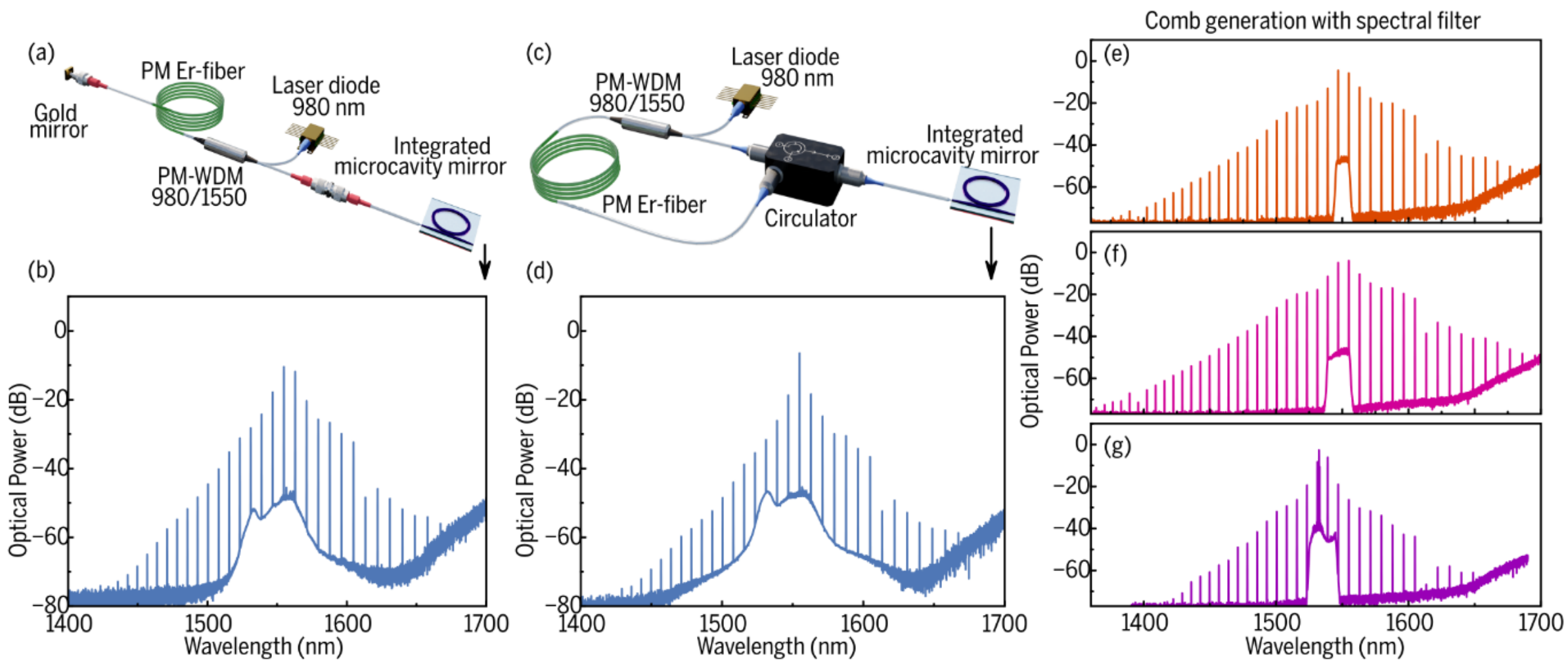

**Figure S7. Variations of hybrid-laser scheme and corresponding comb spectra**. Linear (a) and ring (c) fiber cavity setups with the nested microresonator providing similar soliton self-starting performance. Corresponding comb spectra at 400 mA pump current for the linear scheme (b) and 650 mA pump current for the ring scheme (d). The less efficient comb generation in the ring scheme originates from additional losses on circulator and single pass through the gain media in ring scheme. On the right panel comb spectra are shown at the same pump power (e) with 10 nm spectral filter, (f) spectral filter increased to 18 nm and (g) spectral filter position blue-shifted with respect to central line with ring scheme.

## 3. Numerical model of soliton generation and output pulse measurements

**Numerical model.** We build numerical model using the mean-field formalism introduced for laser-cavity soliton microcombs by Bao H. and co-workers[3]. Their model couples a Kerr microresonator and an external fiber loop with gain medium via discrete supermodes. In our work, we extend that formalism by including Rayleigh back-scattering in the microresonator's equations: a portion of the clockwise (CW) field is scattered into the counter-clockwise (CCW) mode, which then couples into the fiber cavity, amplified and fed back into the microresonator to form the fiber-laser cavity.

The intracavity field in the microring is represented by the slowly varying envelope $a_{cw}(x,t)$ propagating in CW direction and by CCW propagating field $a_{ccw}(x,t)$ induced by Rayleigh back-scattering. Here $x\in[-1,1]$ is the fast time over one round-trip of the microresonator and t is the slow evolution time over many cavity round-trips. The external fiber cavity is described by a set of 2N+1 supermodes $b_q(x,t)$ labelled by an integer index q which represents the offset of each longitudinal mode with respect to the center of microresonator resonance.

In normalized units the governing equations used in the simulations are:

$$\frac{\partial a_{cw}}{\partial t} = \left[-\frac{\kappa_{tot}}{2} - \frac{i\zeta_a}{2}\frac{\partial^2}{\partial x^2} + i|a_{cw}|^2\right] a_{cw} + iRa_{ccw} + \sqrt{\kappa_{ex}}\sum_q b_q, \ (1)$$

$$\frac{\partial a_{ccw}}{\partial t} = \left[-\frac{\kappa_{tot}}{2} - \frac{i\zeta_a}{2}\frac{\partial^2}{\partial x^2} + i|a_{ccw}|^2\right] a_{ccw} + iRa_{cw}, \ (2)$$

$$\frac{\partial b_q}{\partial t} = \left[g + 2\pi i(\Delta - q) + (\sigma - \frac{i\zeta_b}{2})\frac{\partial^2}{\partial x^2}\right] b_q - \sum_q b_q + \sqrt{\kappa_{ex}}a_{ccw}. \ (3)$$

The first two equations (1-2) describe the CW and CCW fields in the microresonator with 0.997 THz FSR. The total loss rate is $\kappa_{tot}$, $\kappa_{ex}$ is the external coupling rate, $\zeta_a$ is the second-order dispersion of the microring, and the cubic term corresponds to the Kerr nonlinearity. Linear coupling R between $a_{cw}$ and $a_{ccw}$ models Rayleigh back-scattering. Equation (3) governs the dynamics of each fiber-cavity supermode. FSR of the fiber laser is ~70 MHz. The parameter σ plays the role of an effective group-delay dispersion and spectral filtering in the external fiber cavity, band-pass filter, and other intracavity components, while Δ sets the detuning between the fiber-cavity modes and the center of microresonator resonance. Each integer q introduces a frequency shift of one fiber-cavity FSR. The field $a_{ccw}$ in the ring drives all supermodes through the coupling term proportional to $\kappa_{ex}$. The gain coefficient is modeled by a constant $g(t)=g_p$. The total energy stored in the fiber cavity at given slow time is $E(t) = \sum_q \int |b_q(x,t)|^2 \, dx$.

Initial conditions consist of a weak continuous-wave background in the CW ring mode $a_{cw}(x,0)$, a localized Gaussian seed in the CCW mode $a_{ccw}(x,0)$ and low-level complex Gaussian noise in all fiber-cavity supermodes $b_q(x,0)$. Initial noise seeding allows the system to evolve into a single soliton state. All parameters used in the model are listed in the Supplementary Table.

Supplementary Table. Parameters used in numerical simulations.

| Parameter | Value (normalized units) | Description |
|---|---|---|
| Total microring loss $\kappa_{tot}$ | 1.94π | Total decay rate of the ring mode |
| External coupling rate $\kappa_{ex}$ | $1.058\kappa_{tot}$ | Linear coupling between the microring and the fiber cavity |
| Rayleigh back-scattering R (CW←→CCW) | $\kappa_{tot}$ | Linear coupling from CW to CCW |
| Ring dispersion $\zeta_a$ | 0.0348 | Second-order dispertion of microresonator |
| Fiber dispersion $\zeta_b$ | 0.59 | Second-order dispertion of the fiber cavity |
| Fiber-cavity dispersion/filtering $\sigma$ | 0.034 | Effective GVD + spectral filtering acting on supermodes ($b_q$) |
| Number of fiber-cavity supermodes N | N=6 → 2N+1=13 supermodes | |
| Gain in fiber cavity $g_p$ | 0.395 | Constant (unsaturated) EDFA gain |
| Detuning $\Delta$ | 0.415 | Relative alignment between center of microresonator resonance and fiber cavity |

**Linearization and modulational-instability (MI) map used to select operating points.** To identify parameter ranges (detuning, gain, dispersion, coupling) that support self-starting solitons, we performed a standard MI analysis by linearizing Eqs. (1)-(3) around the homogeneous ("zero-pattern") state[5]. The Kerr terms vanish to first order at this operating point, so the linear operator contains only linear loss/dispersion, Rayleigh back-scattering, coupling, and loop gain/filtering.

We analyze stability by linearizing the equations and assuming perturbations $e^{i\omega x+\lambda t}$, where spatial frequency $\omega = 2\pi f_x$:

$$\{a_{cw}, a_{ccw}, b_q\}(t,x) = \{\widetilde{a_{\mathrm{cw}}}, \widetilde{a_{\mathrm{ccw}}}, \widetilde{b_q}\}exp(\lambda(\omega)t + i\omega x), (4)$$

and assemble the state vector

$$\boldsymbol{X} = [\widetilde{a_{\mathrm{cw}}}, \widetilde{a_{\mathrm{ccw}}}, \widetilde{b_{-N}} \dots \widetilde{b_N}] \in \mathbb{C}^{(2+M)}, (5)$$

Spatial derivatives act as $\partial \mapsto \omega$ and $\frac{\partial^2}{\partial x^2} \mapsto -\omega^2$. With this convention the eigenproblem reads as

$$\lambda(\omega)\boldsymbol{X} = L(\omega)\boldsymbol{X},\ L(\omega) \in \mathbb{C}^{(2+M)x(2+M)},\ M = 2N+1, (6)$$

and we obtain block matrix: $L(\omega) = \begin{bmatrix} L_{cw} & iR & \sqrt{\kappa_{ex}} \\ iR & L_{ccw} & 0 \\ 0 & \sqrt{\kappa_{ex}} & diag(L_{b_q}) - \sum_q b_q \end{bmatrix}$, (7)

where the diagonal blocks are

$$L_{cw} = -\frac{\kappa_{tot}}{2} - \frac{i\zeta_a}{2}\omega^2, (8)$$

$$L_{ccw} = -\frac{\kappa_{tot}}{2} - \frac{i\zeta_a}{2}\omega^2, (9)$$

$$L_{b_q} = g + 2\pi i(\Delta - q) + (\sigma - \frac{i\zeta_b}{2})\,\omega^2. (10)$$

For each (Δ, g) and each ω we compute the spectrum of L(ω) and define the MI gain as the largest growth rate $G(\omega;\Delta,g) = \max_k \mathrm{Re}\,\lambda_k(\omega;\Delta,g)$. Modulational instability is present if G>0. The frequency of the fastest-growing perturbation is $|f_x(\Delta,g)| = \arg\max_{\omega\geq 0} G(\omega;\Delta,g)$.

**The MI threshold curve** used in the (Δ, g) maps is defined by the sign change of the maximal real part at zero spatial frequency:

$$\max_k \mathrm{Re}\,\lambda_k(\omega = 0;\Delta,g_{thr}) = 0,\ \ g_{thr}(\Delta) = min\,\{g: G(0;\Delta,g) > 0\}. (11\text{-}12)$$

Scanning the (Δ, g) plane produces the MI lobes used to choose operating points for the time-domain simulations (see Figure S3). Their curvature and contrast are set by the ring and loop dispersion $\zeta_{a,b}$ and by the filter parameter σ.

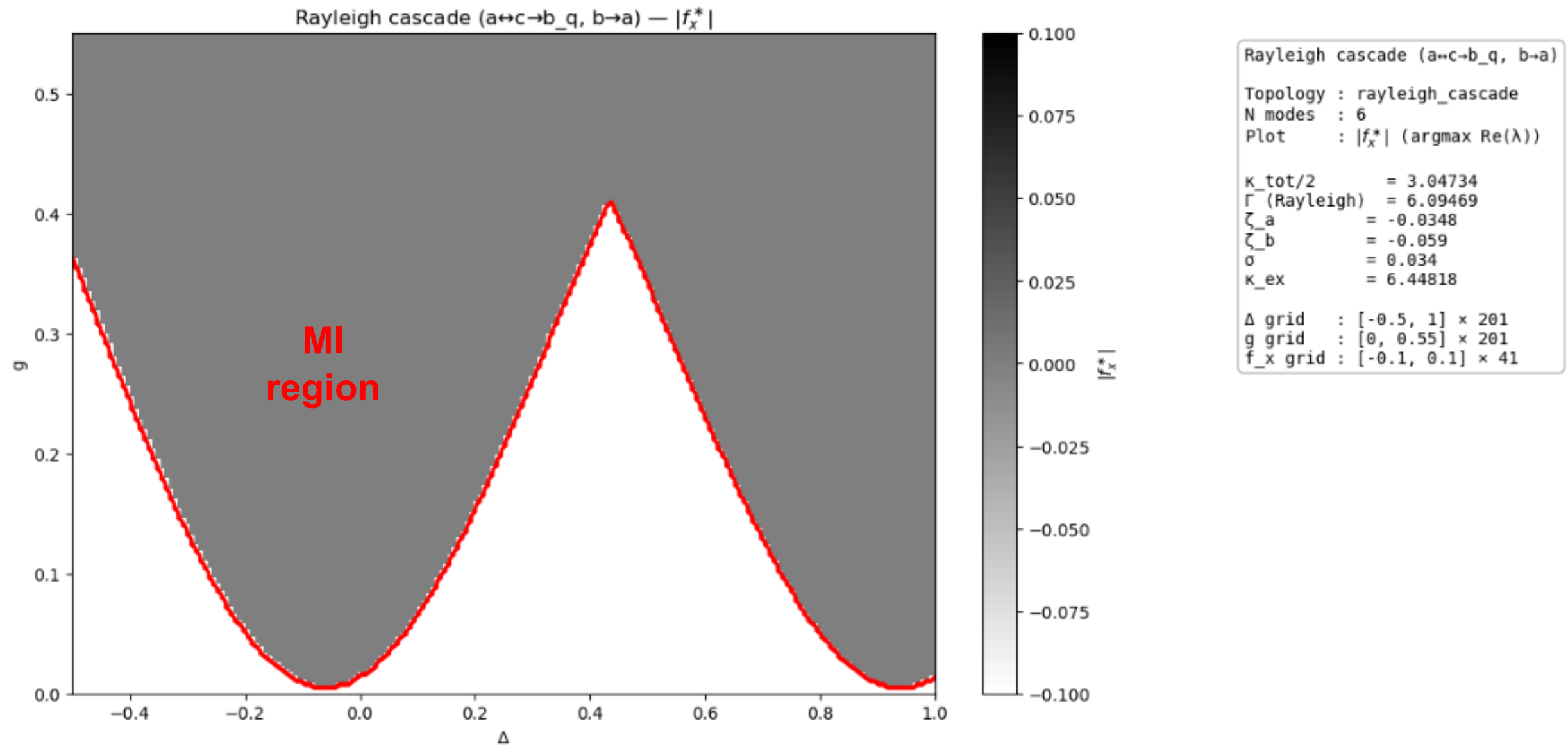

**Figure S3. MI map.** Grayscale shows the spatial frequency of the fastest-growing perturbation, $|f_x|$, obtained from the largest real eigenvalue of the linearized operator. The red curve marks the modulational-instability (MI) boundary where max Re(λ) =0. The MI lobes are centered near Δ≃0 and Δ≃0.5.

**4. Autocorrelation measurement of output pulse train with an external EDFA.** The pulse energy on microresonator output is insufficient for direct autocorrelation (ACF) measurements, so the pulse train is first amplified using an external EDFA. Because the EDFA bandwidth is significantly narrower than the full microcomb spectrum, only a portion of the comb is amplified (see Figure S4a, blue line). In resulting comb three main lines are 15 dB higher than other comb line, this spectral truncation leads to an artificially broadened pulse train (Figure S4b, blue line).

To quantify the effect of the EDFA, we take the simulated microresonator output field and apply a numerical EDFA model with a Gaussian gain profile reproducing the gain spectra width and shape of the amplifier (see Figure S4a, inset). The resulting spectrum matches the experimentally observed spectrum after amplification (Figure S4a, red dashed line), and the autocorrelation of this EDFA-filtered simulated pulse coincides with the measured ACF (Figure S74, red line).

In addition to measuring the EDFA-amplified output, we also record the autocorrelation of the reflected comb and of the pulse train after the active fiber section inside the laser (Figure S4c-d). Although these intracavity traces do not provide the true pulse shape due to their limited spectral bandwidth, they exhibit a clear 1-ps periodicity. This demonstrates that the system supports a pulsed regime not only at the microresonator output but also on the pulses circulating within the fiber cavity and re-entering the microresonator.

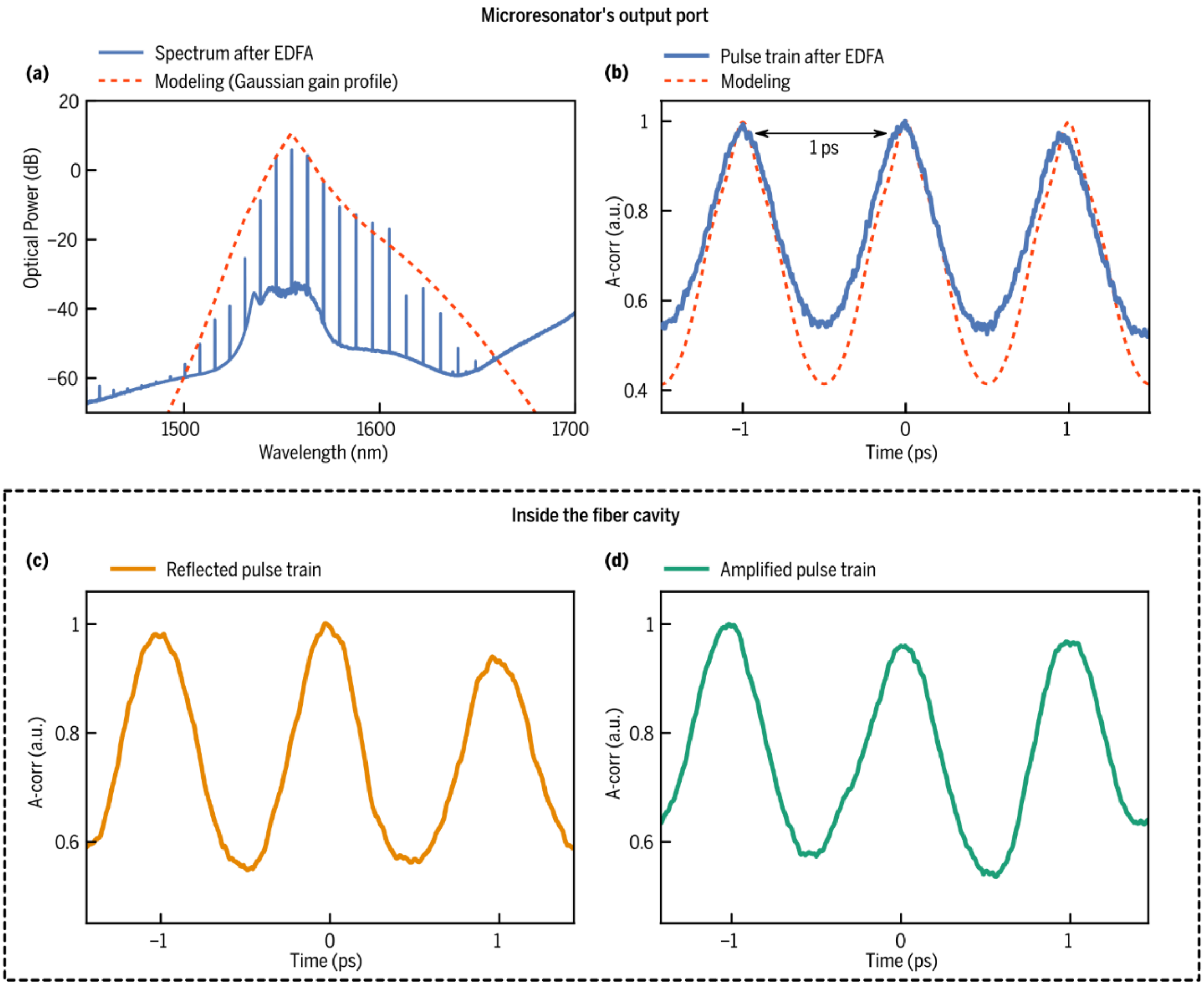

**Figure S4. Autocorrelation measurements at the microresonator's output and inside the fiber cavity.** (a) Spectrum at the microresonator output after external EDFA amplification (blue) together with the simulated spectrum obtained by passing the modeled output field through a numerical EDFA (red dashed). The EDFA bandwidth amplifies only a portion of the full microcomb, resulting in a spectrally truncated pulse. (b) Autocorrelation of the EDFA-amplified pulse train measured experimentally (blue) and the corresponding autocorrelation computed from the EDFA-filtered simulated pulse (red). The agreement shows that the measured ACF to the spectrally truncated pulse rather than to the intracavity soliton. (c) Normalized ACF of the reflected comb inside the cavity, showing a clear 1-ps periodicity. (d) Normalized ACF of the pulse train after the active fiber section in the fiber cavity, which also exhibits the same 1-ps repetition rate.

## 5. Time-domain reconstruction from measured comb spectra.

To assess the consistency between our spectral measurements and the time-domain model, we reconstructed the intracavity pulses by taking the inverse Fourier transform of the measured comb spectra under the assumption of a flat spectral phase, $\varphi(\mu)=0$ for all comb lines, corresponding to the case of a fully coherent comb. For transmitted, reflected, and amplified comb, we resample it on a uniform frequency grid, and then build a spectrum $E(\nu) \propto \sqrt{P(\nu)}$ with zero phase. The temporal waveform E(t) is obtained by the inverse discrete Fourier transform, and the intensity trace $|E(t)|^2$ is normalized to unity peak power. The resulting pulse trains (solid lines in Fig. S5) are plotted together with the corresponding time-domain pulses taken directly from the numerical model (dashed lines).

This comparison shows that, even without including a non-trivial phase, the pulse shapes reconstructed from the spectra closely reproduce the simulated model pulses. In particular, for the transmitted and amplified combs the full-width at half-maximum (FWHM) durations obtained from experimental and simulated spectra are in good agreement (see Fig. S5a, c), and the pedestal structure is also reproduced. For the reflected comb (see Fig. S5b), the reconstructed pulse is somewhat broader than in the model, which is reasonable given that the cavity model does not take into account all experimental effects; nevertheless, the characteristic pedestal is still present. Also, the phase in our experiment was not retrieved and remains unknown, these pulse shapes can be considered as a limiting case for our system under full coherence assumption.

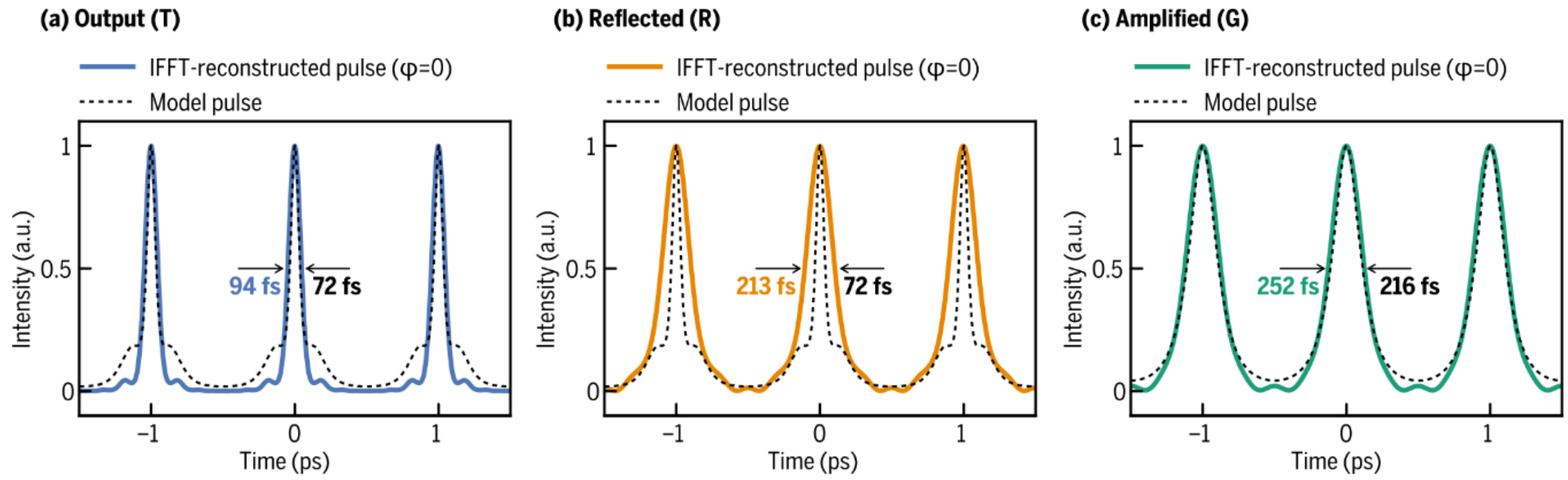

**Figure S5. Temporal pulse reconstruction from comb spectra assuming flat spectral phase, φ(μ)=0.** Solid lines: pulses obtained from the experimental spectra of a) the output, b) reflected, and c) amplified combs. Dashed lines: corresponding time-domain pulses taken directly from the numerical model.

## 6. Laser-scanning spectroscopy measurements

We confirm soliton comb generation by using laser scanning spectroscopy[3,7] to measure the offset (Δ) of the comb lines relative to the MRR resonance center and we analyzed the beating signal between the external CW laser and the microcomb lines (**Figure S6a**). Generally, blue-detuned states correspond to a positive frequency offset (Δ>0), while red-detuned states correspond to a negative offset (Δ<0) indicating that the system operates in the soliton regime, where the Kerr nonlinearity balances the cavity dispersion, enabling stable soliton formation [8,9].

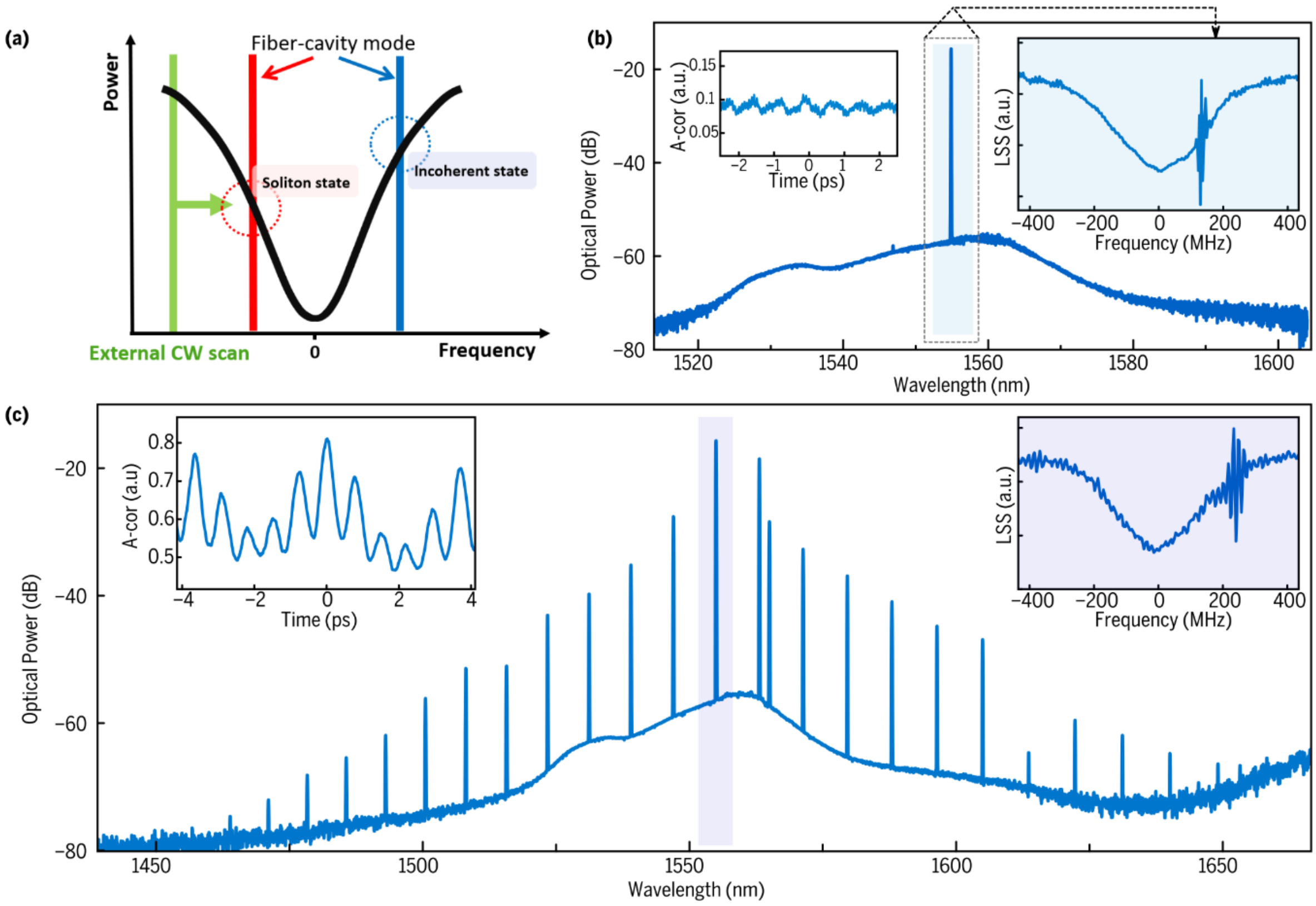

**Figure S6. Laser-scanning spectroscopy technique**. (a) Schematic representation of the spectral distribution of a cold MRR resonance (black curve) with fiber-cavity modes depicting different offsets relative to the microcavity resonance center. The red curve represents the red-detuned soliton state, the blue curve represents the blue-detuned incoherent state, and the green line indicates the external tunable CW laser. Beating signals arise at the dotted points where the CW laser scan intersects the fiber-cavity mode within the MRR resonance. (b) Output comb spectra and their corresponding autocorrelation traces (top-left inset) and laser-scanning spectroscopy (LSS) traces (bottom-right inset). (c) Spectral measurements for both soliton and incoherent states. Insets illustrate autocorrelation traces (left) and LSS results (right). The LSS traces validate transitions between coherent soliton states and incoherent states.

To perform these measurements, an external tunable CW laser with a 10 kHz linewidth was introduced into the fiber cavity via a fiber optic circulator connected to a 70/30 coupler (**Figure S7**). The power of the CW laser entering the MRR did not exceed 1 mW, ensuring that it did not affect the shape of the hot resonance. The CW laser frequency was swept across one of the MRR resonances during comb generation, demonstrating a beating when coincided with a comb line. Simultaneously, the output characteristics were measured using a spectrometer, a power meter, an oscilloscope, and an autocorrelator.

In addition to the soliton combs with red-detuning discussed in the main text, by tuning the position of the lensed fiber probe we also observed comb and CW states with blue-detuning illustrated in Figure S6b-c, which exhibit distinct spectral and temporal characteristics compared to the soliton regime. Cavity length and losses play a critical role in determining the system's operating regime and enabling transitions between soliton states and noncoherent states.

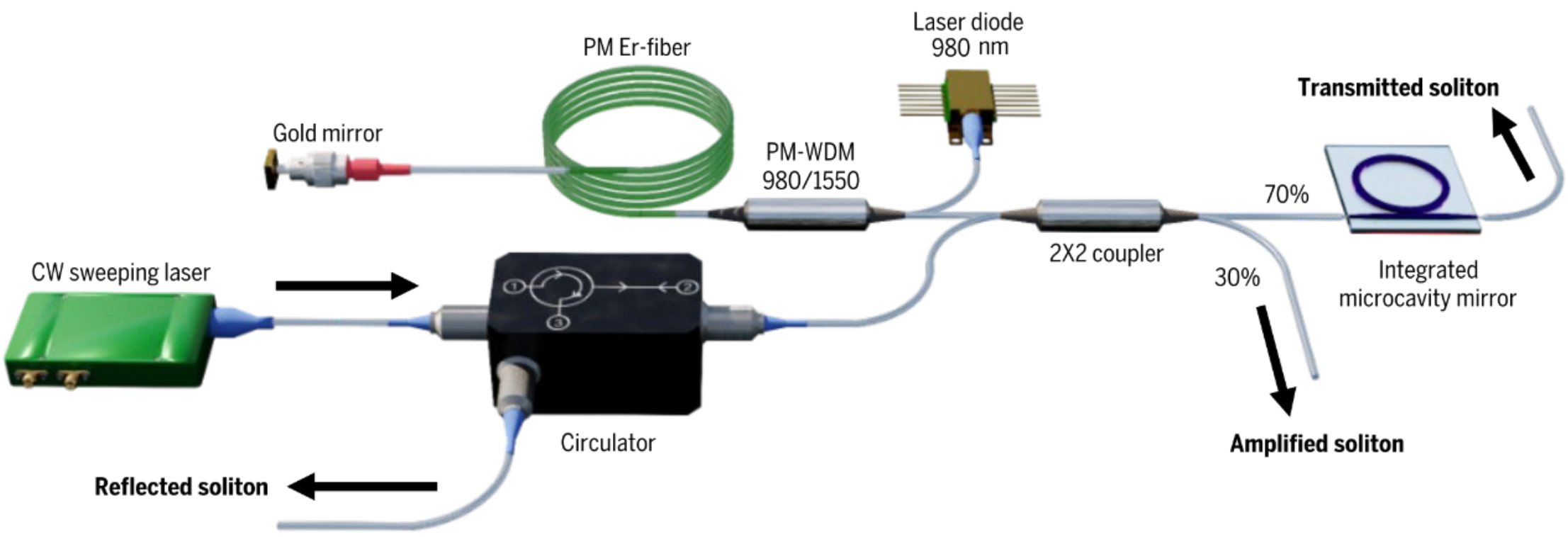

**Figure S7. Setup for investigating intracavity comb power, spectral dynamics, and laser-scanning spectroscopy.** The experimental setup comprises a nonlinear MRR functioning as a semitransparent mirror, an erbium-doped fiber amplifier (EDFA), and a gold mirror providing optical feedback. A 2×2 optical coupler was integrated to simultaneously measure intracavity characteristics and perform laser scanning spectroscopy by sweeping the CW laser across the MRR resonances. This configuration enables the detection of the operating laser mode's position relative to the MRR resonance.

## 7. Reflection and Transmission characteristics.

**In Figure S8a** we demonstrate reflected and transmitted power fractions as measured in the fiber as a function of incident power from fiber to chip. The orange line corresponds to the total useful power (R+T) and stays around 12% regardless of the pump. The remaining losses (88%) come from the losses in the chip and coupling. To prove that nonlinear behavior of reflected and transmitted power from MRR on the pump is an intrinsic property of MRR-laser hybrid system and cannot be referred, for example, to variations of the coupling to the chip induced by the heating, we measured reflection and transmission as a function of incident power using external sources (ASE source and CW laser) that did not form a feedback system with the microresonator (Figure S8b-c). As expected, in these cases, the reflection and transmission remained constant and did not exhibit power-dependent changes. The reflection of approximately 1%, we attribute to the reflection from the chip facet.

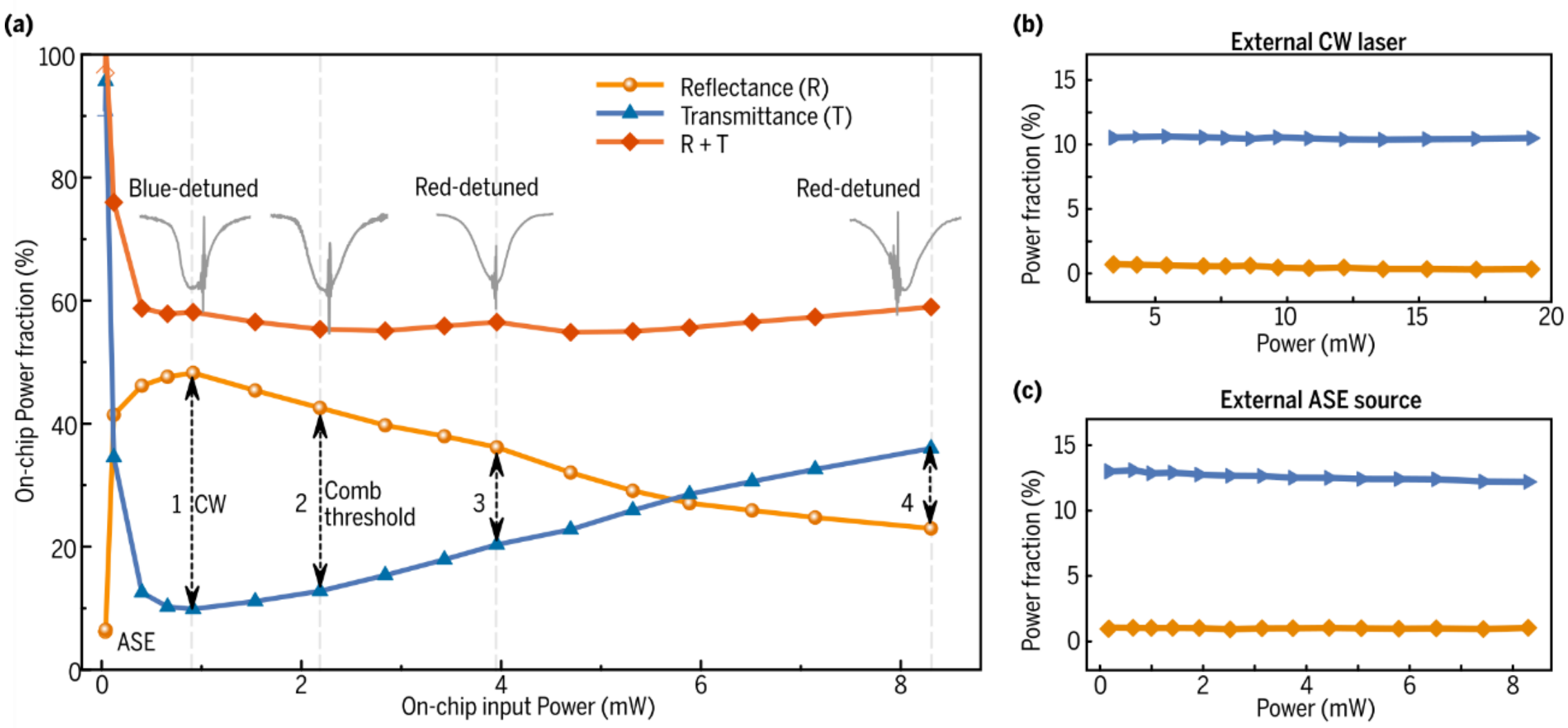

**Figure S8. Measured reflection and transmission power fractions in comb generation regime and under external light sources.** a) Reflection (red line), transmission (blue line), and their sum (orange line) are shown as a function of the fiber intracavity power after amplifier. The inset shows the evolution of the mode detuning for different regimes: blue-detuned CW state and red-detuned comb state. Power fractions of reflected (red) and transmitted (blue) light measured in the fiber using an external CW laser (b) and ASE source (c).